\documentclass[preprint, twocolumn]{aastex631}

\usepackage{graphicx}
\usepackage{amsmath}	
\usepackage{amssymb}
\usepackage{upgreek}
\usepackage{bm}
\usepackage{subfigure}

\submitjournal{AAS Journal}

\shorttitle{A new approach to constraining properties of AGN host galaxies}
\shortauthors{Yu et al.}

\begin{document}

\title{A new approach to constraining properties of AGN host galaxies by combining image and SED decomposition: testing upon the $M_{\rm BH}-M_\star$ relation}


\correspondingauthor{Lulu Fan,Yunkun Han}
\email{llfan@ustc.edu.cn,hanyk@ynao.ac.cn}

\author[0009-0008-1319-498X]{Haoran Yu}
\affiliation{CAS Key Laboratory for Research in Galaxies and Cosmology, Department of Astronomy,
University of Science and Technology of China, Hefei 230026, China}
\affiliation{School of Astronomy and Space Science, University of Science and Technology of China, Hefei 230026, China}

\author[0000-0003-4200-4432]{Lulu Fan}
\affiliation{CAS Key Laboratory for Research in Galaxies and Cosmology, Department of Astronomy,
University of Science and Technology of China, Hefei 230026, China}
\affiliation{School of Astronomy and Space Science, University of Science and Technology of China, Hefei 230026, China}
\affiliation{Deep Space Exploration Laboratory, Hefei 230088, China}

\author[0000-0002-2547-0434]{Yunkun Han} 
\affiliation{Yunnan Observatories, Chinese Academy of Sciences, 396 Yangfangwang, Guandu District, Kunming, 650216, P. R. China}
\affiliation{Center for Astronomical Mega-Science, Chinese Academy of Sciences, 20A Datun Road, Chaoyang District, Beijing, 100012, P. R. China}
\affiliation{Key Laboratory for the Structure and Evolution of Celestial Objects, Chinese Academy of Sciences, 396 Yangfangwang, Guandu District, Kunming, 650216, P. R. China}
\affiliation{International Centre of Supernovae, Yunnan Key Laboratory, Kunming 650216, P. R. China}

\author[0009-0004-7885-5882]{Weibin Sun}
\affiliation{CAS Key Laboratory for Research in Galaxies and Cosmology, Department of Astronomy,
University of Science and Technology of China, Hefei 230026, China}
\affiliation{School of Astronomy and Space Science, University of Science and Technology of China, Hefei 230026, China}

\author[0009-0003-9423-2397]{Yihang Zhang}
\affiliation{CAS Key Laboratory for Research in Galaxies and Cosmology, Department of Astronomy,
University of Science and Technology of China, Hefei 230026, China}
\affiliation{School of Astronomy and Space Science, University of Science and Technology of China, Hefei 230026, China}

\author[0000-0001-8917-2148]{Xuheng Ding}
\affiliation{School of Physics and Technology, Wuhan University, Wuhan 430072, China}

\author[0000-0002-1935-8104]{Yongquan Xue}
\affiliation{CAS Key Laboratory for Research in Galaxies and Cosmology, Department of Astronomy,
University of Science and Technology of China, Hefei 230026, China}
\affiliation{School of Astronomy and Space Science, University of Science and Technology of China, Hefei 230026, China}



\begin{abstract}

The outshining light from active galactic nuclei (AGNs) poses significant challenges in studying the properties of AGN host galaxies.
To address this issue, we propose a novel approach which combines image decomposition and spectral energy distribution (SED) decomposition to constrain properties of AGN host galaxies.
Image decomposition allows us to disentangle optical flux into AGN and stellar components, thereby providing additional constraints on the SED models to derive more refined stellar mass.
To test the viability of this approach, we obtained a sample of 24 X-ray selected type I AGNs with redshifts ranging from 0.73 to 2.47.
We estimated the stellar masses for our sample and found that our results are generally consistent with earlier estimates based on different methods.
Through examining the posterior distribution of stellar masses, we find that our method could derive better constrained results compared to previous SED decomposition methods.
With the derived stellar masses, we further studied the $M_{\rm BH}-M_\star$ relation of our sample, finding a higher intrinsic scatter in the correlation for our entire sample compared to the local quiescent correlation, which could be caused by a few ``black hole monsters'' in our sample.
We propose that based on our method, future works could extend to larger samples of high-redshift AGN host galaxies, thereby enhancing our understanding of their properties.

\end{abstract}



\section{Introduction} \label{sec:intro}

It is widely believed that supermassive black holes (SMBHs) reside in the central regions of most galaxies \citep[e.g.,][]{magorrian1998, gebhardt2001, kormendy2013}.
Moreover, tight correlations between BH mass ($M_{\rm BH}$) and the properties of the bulge component of the host galaxy (e.g., $M_{\rm BH}-M_{\rm bulge}$ relation) have been found in the local universe \citep{ferrarese2000, marconi2003, haring2004, bennert2010, beifiori2012, schramm2013}. 
When it comes to higher redshifts, separating bulge mass from disk mass becomes difficult, making it harder to study the correlation between BH mass and total stellar mass of the host galaxy \citep[e.g.,][]{ding2020, suh2020, li2021a, li2024, zhuang2023, tanaka2024}.

In order to explain the origin of this correlation, various theoretical models have been proposed.
For example, AGN feedback is invoked in many cosmological simulations, in which a fraction of AGN energy heats the surrounding gas and regulates the BH accretion and the star formation of the host galaxy \citep{springel2005, dimatteo2008, hopkins2008, degraf2015, harrison2017}.
Alternatively, introducing an indirect connection in which BH accretion and star formation share a common gas supply is viable \citep{cen2015, menci2016, angles-alcazar2017}.
On the other hand, in a cosmic averaging scenario the correlations could be established through mergers based solely on Central Limit Theorem, without involving any close physical connection between the BH and the host galaxy \citep{peng2007, hirschmann2010, jahnke2011}.
However, which of these models is the dominant cause of the mass correlation is still in debate.

In order to investigate the potential coevolution, many observational efforts have been put into studying the correlations between BHs and host galaxies for moderate and high redshifts \citep[e.g.,][]{treu2004, treu2007, jahnke2009, schramm2013, sun2015, park2015, suh2020, ding2020, li2021a, zhuang2023, tanaka2024}. 
However, obtaining either the BH mass or the properties of the host galaxies is fraught with difficulties.
The BH mass can be obtained with the dynamical method for nearby galaxies, but it becomes difficult with higher redshifts because of the demand of high spatial resolution. 
Although recently BH mass has been dynamically measured for a quasar with $z\sim2$ \citep{abuter2024}, previous high redshift BH masses are derived mainly by the virial method \citep[e.g.,][]{vestergaard2006, schulze2018} for type I AGNs. 
Without the obscuration of the dusty torus according to the unified AGN model \citep{netzer2015}, the accretion disk in a type I AGN glares and outshines the stellar emission of the host galaxy by far in the optical and ultraviolet (UV) bands, bringing difficulty to separating the stellar component from AGN contamination.

However, there have been several methods to estimate the host stellar masses of these unobscured AGNs.
For example, \citet{matsuoka2015} obtained high signal-to-noise ratio (S/N) optical spectra of 191 broad-line quasars with $z < 1$ and performed spectral decomposition to derive the velocity dispersion and the mass of the stellar component.
UV-to-FIR (far-infrared) broadband SED decomposition \citep{zou2019, suh2019, suh2020, sun2024} was employed to derive stellar properties of AGN host galaxies, mainly by adding an AGN template in the SED models.
With recent deep imaging surveys, image decomposition techniques have been employed to disentangle host galaxies from core contamination \citep{peng2002, park2015, kim2017, ding2020, ding2023, li2021a, li2021, li2023, li2024, zhuang2023, zhuang2024}. 
In the image decomposition process, the AGN contribution is modeled as a point spread function (PSF) and is subtracted from the original image, leaving the emission from the host galaxies.
\citet{ding2020} estimated stellar masses of 32 X-ray-selected AGNs based on stellar fluxes obtained through image decomposition. 
\citet{setoguchi2023} used image decomposed SED of the stellar components derived by \citet{li2021a} to mitigate contamination in determining the BH mass.
Inspired by the aforementioned works, here we propose a novel method combining image and SED decomposition to better constrain the stellar mass of host galaxies, where we use the fluxes of the AGNs and the host galaxies derived from image decomposition to refine the corresponding models in the UV-FIR broadband SED fitting.

This paper is structured as follows.
Section \ref{sec:sample} introduces the sample used in this work.
Section \ref{sec:method} describes the Bayesian methodology that we utilized to incorporate additional information from image decomposition to constrain SED models.
Section \ref{sec:results} compares our stellar mass estimation with previous work and examines the $M_{\rm BH}-M_\star$ relation for our sample.
Section \ref{sec:discussion} discusses the probable sources of errors in our work and the future prospects of our approach.
Finally, Section \ref{sec:summary} presents the concluding remarks.
Throughout the paper, we assume a flat $\mathrm{\Lambda CDM}$ cosmology \citep{komatsu2011} with $\Omega_\Lambda=0.7$, $\Omega_{\rm m}=0.3$ and $H_0=70\ {\rm km\,s^{-1}\,Mpc^{-1}}$. The luminosities are given in units of $\rm erg\, s^{-1}$ and the masses are given in units of $\rm M_\odot$.

\section{Sample and data} \label{sec:sample}

We initially selected a sample of moderate-luminosity type I AGNs from \citet{schulze2018}, hereafter S18, in which the near-infrared spectra of the AGN broad line region (BLR) were measured and the black hole masses and spectroscopic redshifts were determined.
The sources in our sample are distributed in the CANDELS COSMOS, UDS, and GOODS-S fields, which makes it easier to search for legacy multiband data.
For AGNs with black hole mass derived from H$\alpha$ or H$\beta$ lines \citep{trakhtenbrot2012}, we searched for available \textit{HST} multi-band imaging data and managed to obtain the UV-FIR broadband photometric data for these AGNs from archival catalogs.

\begin{figure}[htbp]
    \centering
    \includegraphics[width=\columnwidth]{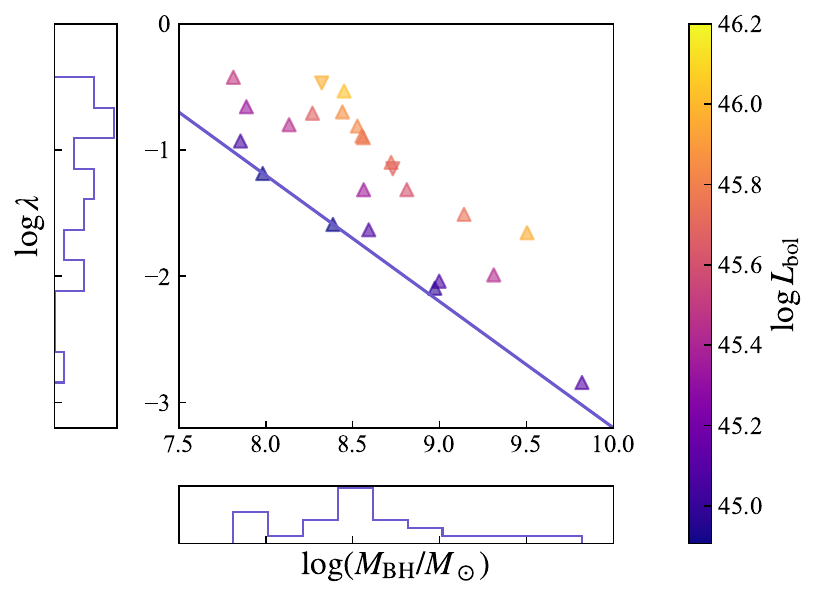}
    \caption{Distributions of $\log (M_{BH}/M_\odot)$, $\log \lambda$ and $\log L_{\rm bol}$ of the AGNs selected from S18.
    In the central panel, our samples are plotted in color-mapped triangles.
        The upward triangles indicate samples using H$\rm \alpha$ line while the downward ones indicate samples using H$\rm \beta$ line.
    The blue line represents the sources with $\log L_{\rm bol}(\rm erg ~ s^{-1})=44.9$, which equals to the lower $L_{\rm bol}$ limit of our sample.
    The left and the bottom panels show the marginal distributions of the two parameters, respectively.
    The color bar on the right maps the bolometric luminosity of the sample.}
    \label{fig:sample}
\end{figure}

We retrieved multi-band \textit{HST} WFC3/ACS images from the 3D-HST survey \citep{momcheva2016}.
Through crossmatching the S18 catalog and Hubble Source Catalog \citep{whitmore2016}, we further obtained archival imaging data from the Mikulski Archive for Space Telescopes (MAST) at the Space Telescope Science Institute\footnote{The specific observations analyzed can be accessed via \dataset[doi: 10.17909/5yh3-w824]{https://doi.org/10.17909/5yh3-w824}.}.
We obtained UV-FIR broadband photometric data for our sample from various archival catalogs, namely \textit{GALEX} \citep{martin2005} for UV, \textit{SDSS} \citep{ahumada2020} and \textit{PAN-STARRS} \citep{chambers2016} for optical, \textit{HST} \citep{whitmore2016} for optical/near-infrared (NIR), \textit{UKIRT} \citep{lawrence2007, hambly2008} and \textit{VISTA} \citep{cross2012} for NIR, and \textit{WISE} \citep{cutri2021}, \textit{Spitzer} \citep{LeFloch2009, sanders2007} and \textit{Herschel} \citep{griffin2010, pilbratt2010} for mid-infrared (MIR) and FIR. 
We set the matching radius at 1 arcsecond for UV and optical bands.
For MIR and FIR bands, given the poorer spatial resolution and pointing precision, we extended the matching radius to 5 arcseconds.
Furthermore, we visually inspected the corresponding imaging data and removed the photometric data likely to be contaminated by nearby bright sources.

We employed image decomposition for the matched sources and visually checked the decomposition results.
Given the limited image quality of the archival data, sources with poorly constrained PSFs
tend to yield significantly biased decomposition results and were therefore excluded from further investigation.
Our final sample consisted of 24 type I AGNs, with redshift spanning $0.72<z<2.47$, black hole mass $7.81<\log M_{\rm BH}(\rm M_\odot)<9.82$, Eddington ratio $-2.84<\log \lambda <-0.42$ and inferred bolometric luminosity 
$44.91 <\log L_{\rm bol}({\rm erg\;s^{-1}})<46.02$. 
Figure \ref{fig:sample} shows the properties of the selected BHs, which helps us correct for the selection bias.

\section{Method} \label{sec:method}

In this section, we illustrate our new method which combines image decomposition and SED decomposition to help to reduce the degeneracy between AGN bright UV-optical emission and host stellar emission.
First, the image decomposition procedure is briefly introduced in Section \ref{sec:decomp}.
Then Section \ref{sec:sed} describes how we use image decomposition results as extra constraints in the UV-FIR broadband SED decomposition.

\subsection{\textit{HST} Image decomposition} \label{sec:decomp}

Following common practice, we simultaneously fitted the two-dimensional flux distribution of the central AGN and the underlying host galaxy, which were modeled as an unresolved point-source component characterized by the PSF and a two-dimensional Sersic profile \citep[e.g.,][]{peng2002, park2015, kim2017, ding2020, li2021a}.
Although a single Sersic profile may cause uneven residuals to be reported when substructures such as spiral arms are present, it acts as an adequate first-order approximation to estimate the total flux of the stellar component, which has been reported in the literature \citep{matsuoka2015, sun2015, suh2020, ding2020, li2021a}.

We used a state-of-the-art image modeling tool \textsc{Galight} \citep{ding2021} to perform the AGN-host galaxy image decomposition.
\textsc{Galight} is a Python-based open-source package designed for two-dimensional model fitting of especially large samples of extragalactic sources.
It makes use of the image modeling capabilities of \textsc{Lenstronomy} \citep{birrer2018}, while adding automatic features such as searching for PSF-stars in the field of view, estimating the noise map of the data and identifying objects to set the initial model and associated parameters. 

\begin{figure*}[htb]
    \centering
    \includegraphics[width=\textwidth]{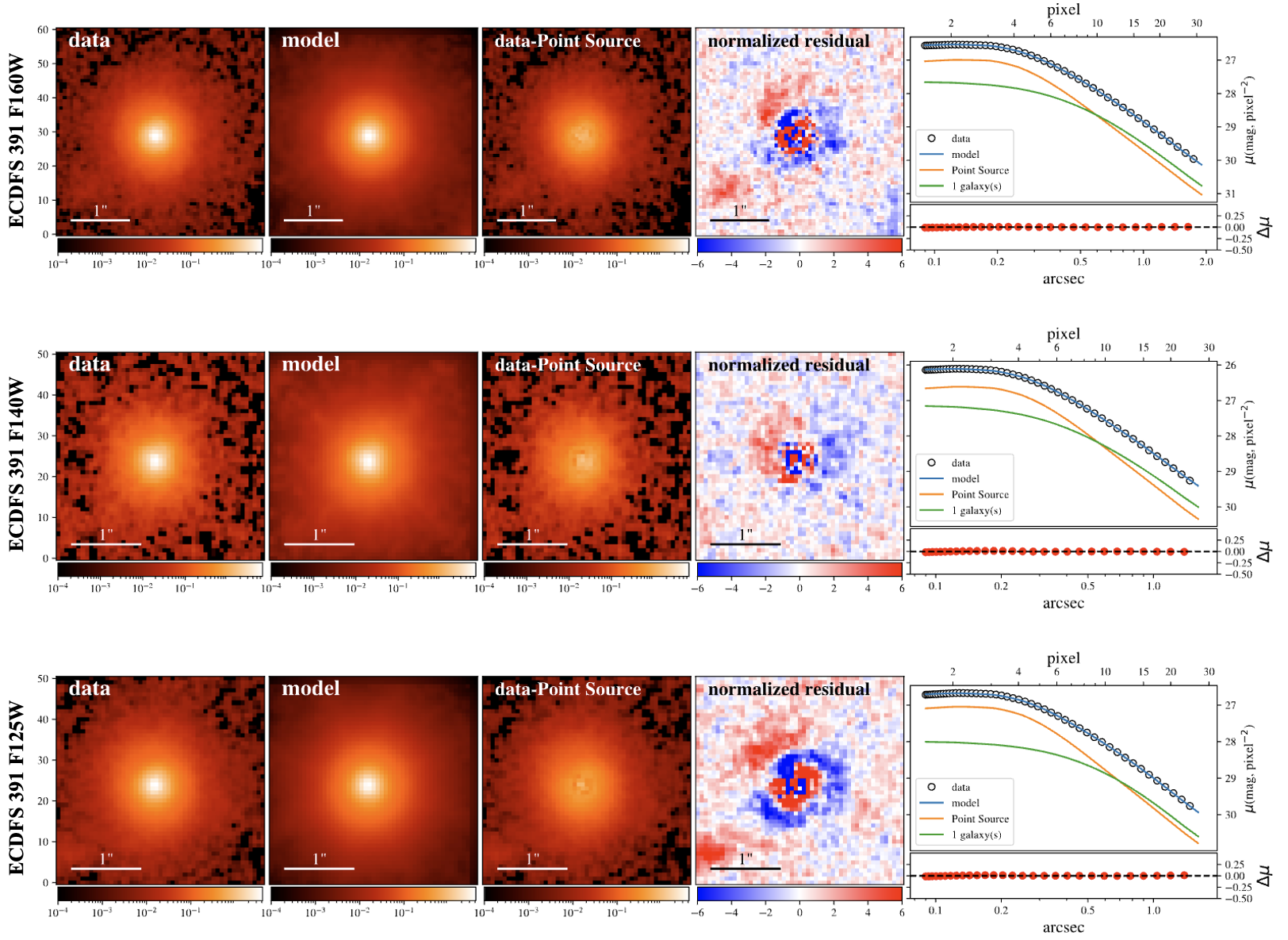}
    \caption{Best image decomposition results for ECDFS 391 in F125W, F140W and F160W of \textit{HST} WFC3. The panels from left to right are: (1) observed data, (2) best-fit Sersic + point source model, (3) observed data minus the point source model (i.e., host galaxy and the simultaneously modeled nearby galaxies), (4) residual divided by variance and (5) radial surface brightness profile (top) and residual (bottom). This profile includes the data (open circles), best-fit model (blue curve), the point source model of AGN (orange curve) and the model of 2 galaxies (green curve). The fitting is based on 2-dimensional image, while the 1-dimensional profile is only an illustration of the fitting result.}
    \label{fig:decomposition}
\end{figure*}

It has been emphasized that PSF is the dominant source of error in the flux decomposition procedure \citep{ding2020, tanaka2024}.
To minimize the potential effect of PSF mismatch, we adopted a weighted method from \citet{ding2020}.
    With the aid of \textsc{Galight}'s automatic PSF searching function, we found isolated and unsaturated point sources within the field of view of each target\footnote{In the processing procedure for a target source, only the PSFs within the same band, near the target, are involved.}.
After visual inspection the profiles with poor SNR are excluded, then we selected the point sources with the top 5 sharpest profiles (i.e. the lowest full width at half maximum) as PSFs to be used in the image decomposing routine.
We also included stacked PSFs in the fitting.
Using multiple PSFs, we can estimate the uncertainty of the derived fluxes.
For a source in a certain band, the weight of each PSF is calculated as
\begin{equation}\label{eq:3}
    w_i = \exp\left( -\alpha \frac{\chi_i^2-\chi_{\rm best}^2}{2 \chi_{\rm best}^2} \right),
\end{equation}
where $i$ is the rank ordered by $\chi^2$ from smallest to largest, $\chi_{\rm best}=\chi_1$ represents the PSF with the best performance, and $\alpha$ is an inflation parameter which satisfies the following equation:

\begin{equation}\label{eq:4}
    \alpha \frac{\chi_{\rm worst}^2 - \chi_{\rm best}^2}{2\chi_{\rm best}^2} = 2.
\end{equation}

The purpose of introducing $\alpha$ is to avoid too little difference between the relative likelihood of different PSFs.
Based on these weights we can calculate the weighted arithmetic mean and the corresponding root mean squared deviation (RMSD) as
\begin{equation}\label{eq:5}
    \bar{x} = \frac{\sum_{i=1}^N x_i\cdot w_i}{\sum_{i=1}^N w_i},
\end{equation}
\begin{equation}\label{eq:6}
    \sigma = \sqrt{\frac{\sum_{i=1}^N (x_i-\bar{x})^2 \cdot w_i}{\sum_{i=1}^N w_i}},
\end{equation}
where $N$ is the quantity of PSFs used for the image of this band.
For those with less than four available PSFs, we fix the RMSD as $10\%$ of the mean value as an estimation. Figure \ref{fig:decomposition} shows the decomposition results for ECDFS 391 in three bands. 
The decomposed fluxes of the host galaxy and the AGN component are shown in Tables \ref{tab:host} and \ref{tab:ps} in the appendix.

\subsection{SED fitting} \label{sec:sed}

By assuming a Gaussian form of noise for all bands, the likelihood function for Bayesian multi-component SED decomposition of galaxy is normally defined as:
\begin{tiny}
\begin{align}
    \mathcal{L}(\bm \theta )&\equiv p(\bm d|\bm \theta, \bm M,\bm I)\notag\\
                            &=p({\bm F_{\rm o}^{\rm TOT}},{\bm F_{\rm o}^{\rm GAL}},{\bm F_{\rm o}^{\rm AGN}}|\bm \theta_1,\bm \theta_2,  M_{\rm GAL}, M_{\rm AGN},\bm I)\notag\\
                            &= \prod\limits_{i = 1}^{n} {\frac{1}{{\sqrt {2\pi } {\sigma_i}}}\exp \left( - \frac{1}{2}\frac{{{{({F_{\rm o,i}^{\rm TOT}} - {s_1*f_{\rm m,i}^{\rm GAL(\bm \theta_1 )}} - {s_2*f_{\rm m,i}^{\rm AGN(\bm \theta_2 )}})}^2}}}{\sigma_i^2}\right)},
                            \label{eq:likelihood_1}
\end{align}
\end{tiny}
where $F_{\rm o,i}^{\rm TOT}$ and  $\sigma_i$ represent the total observational flux and associated uncertainty in each band, 
$f_{\rm m,i}^{\rm GAL(\bm \theta_1 )}$ ($f_{\rm m,i}^{\rm AGN(\bm \theta_2 )}$) represents the flux of the $i$-th band predicted by the galaxy (AGN) SED model, $\bm \theta_1$ and $s_1$ ($\bm \theta_2$ and $s_2$) represent the set of free parameters and scaling factor for the galaxy (AGN) component, $M_{\rm GAL}$ ($M_{\rm AGN}$) represents assumptions involved in the SED modeling of galaxy (AGN), while $\bm I$ represents all other relevant background assumptions.

To take advantage of the additional information from image decomposition of host and AGN in $m$ of $n$ bands, we define the likelihood function as:
\begin{tiny}
\begin{align}
    \mathcal{L}(\bm \theta )&\equiv p(\bm d|\bm \theta, \bm M,\bm I)\notag\\
                            &=p({\bm F_{\rm o}^{\rm TOT}},{\bm F_{\rm o}^{\rm GAL}},{\bm F_{\rm o}^{\rm AGN}}|\bm \theta_1,\bm \theta_2,  M_{\rm GAL}, M_{\rm AGN},\bm I)\notag\\
                            &= \prod\limits_{i = 1}^{n} {\frac{1}{{\sqrt {2\pi } {\sigma_i}}}\exp \left( - \frac{1}{2}\frac{{{{({F_{\rm o,i}^{\rm TOT}} - {s_1*f_{\rm m,i}^{\rm GAL(\bm \theta_1 )}} - {s_2*f_{\rm m,i}^{\rm AGN(\bm \theta_2 )}})}^2}}}{\sigma_i^2}\right)}\notag\\
                            &* \prod\limits_{j = 1}^{m} {\frac{1}{{\sqrt {2\pi } {\sigma_j}}}\exp \left( - \frac{1}{2}\frac{{{{({F_{\rm o,j}^{\rm GAL}} - {s_1*f_{\rm m,j}^{\rm GAL(\bm \theta_1 )}})}^2}}}{\sigma_j^2}\right)}\notag\\
                            &* \prod\limits_{k = 1}^{m} {\frac{1}{{\sqrt {2\pi } {\sigma_k}}}\exp \left( - \frac{1}{2}\frac{{{{({F_{\rm o,k}^{\rm AGN}} - {s_2*f_{\rm m,k}^{\rm AGN(\bm \theta_2 )}})}^2}}}{\sigma_k^2}\right)},
                            \label{eq:likelihood_2}
\end{align}
\end{tiny}
where $F_{\rm o,j}^{\rm GAL}$ and  $\sigma_j$ ($F_{\rm o,k}^{\rm AGN}$ and  $\sigma_k$) represent the observational flux and associated uncertainty of galaxy (AGN) component in each of the $m$ bands derived from AGN-host image decomposition.
This has been implemented in the new version of the BayeSED code \citep{han2014,han2018,han2023}.
Compared to the normally defined likelihood in Equation \ref{eq:likelihood_1}, our likelihood additionally constrains the galaxy (AGN) model with the flux obtained through image decomposition.

\begin{figure*}[htb]
\centering
\includegraphics[width=\textwidth]{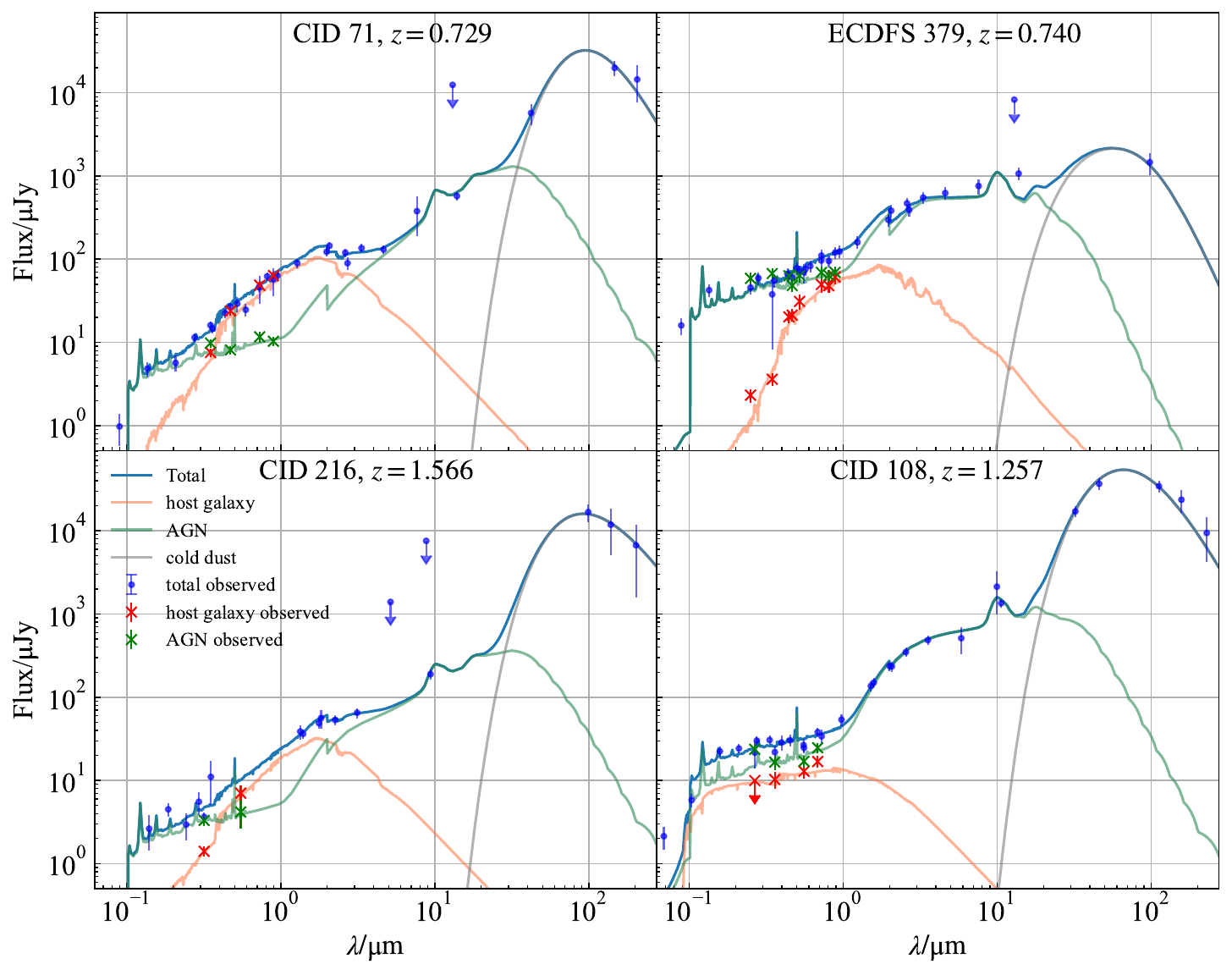}
\caption{Three-component SED decomposition results of 4 sources.
The spectra have been shifted to rest frame according to the spectroscopic redshifts of the sources.
The legend is shown only in the lower left panel.
The orange curves are the stellar population models, assuming star forming history to be double power law.
The green curves are the AGN models, which are composed of a clumpy torus model and an empirical type I quasar template.
The gray curves are the models of the cold dust component.
The blue dots with error bars are the observed UV-FIR fluxes, while the red and green diagnal crosses are the fluxes of stellar component and AGN inferred from image decomposition.
Points with downward arrows denote 3$\sigma$ upper limits.}
\label{fig:sed}
\end{figure*}

For the SED modeling of the stellar population, we employed the \cite{bruzual2003} simple stellar population (SSP) model with a \cite{chabrier2003} stellar initial mass function (IMF), the double power law form of star formation history (SFH),
and the \cite{calzetti2000} dust attenuation law.
As in \cite{han2023}, we additionally employ a linear SFH-to-metallicity mapping model to describe the chemical evolution history of the galaxy, which has been found to increase the Bayesian evidence of the SED model. 
In addition, a cold dust component, which is modeled as a gray body, was added with the energy balance assumption, in which the total energy of stellar emission absorbed by cold dust is equal to that reemitted in the IR.
Finally, the SED of AGN is modelled as the combination of a clumpy torus model and an empirical SED of type I quasar where an SMC-like dust attenuation law has been applied.

Figure \ref{fig:sed} presents 4 representative SED fitting results, demonstrating various scenarios for decomposing AGNs and host galaxies.
A user-friendly documentation \footnote{\url{https://github.com/hanyk/BayeSED3/blob/main/observation/agn_host_decomp/demo.ipynb}} is available on 
the homepage of \texttt{BayeSED}, where readers can access the multi-component photometric data to reproduce our SED fitting results.

\section{Results} \label{sec:results}

\subsection{Stellar mass estimation} \label{subsec:mass}

\begin{figure}[htb]
    \centering
    \includegraphics[width=\columnwidth]{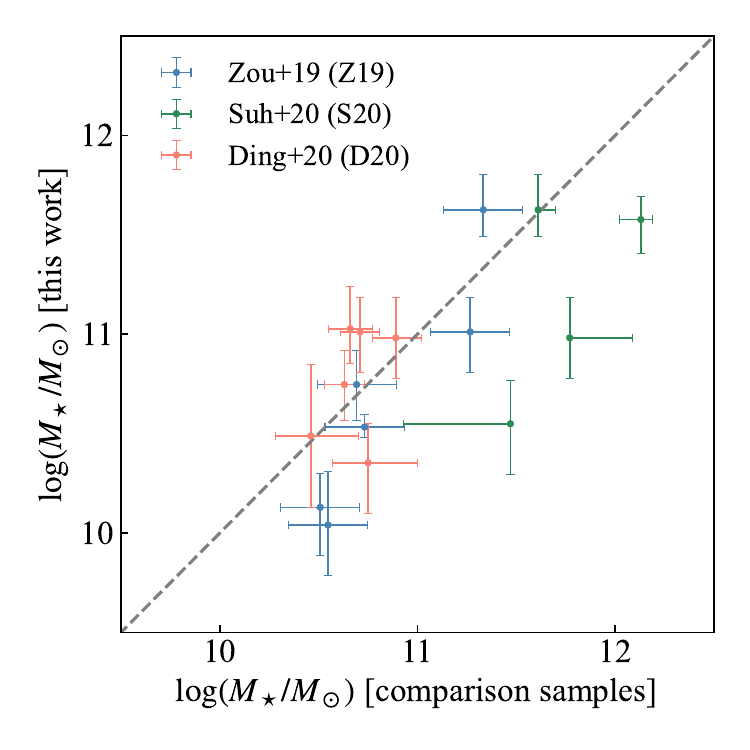}
    \caption{The estimates of our host galaxy stellar masses in comparison with Z19, S20 and D20.
    The $y$-values of the points refer to the stellar masses that we derived, while the $x$-values indicate the stellar masses given by Z19 (S20, D20) for the blue (green, orange) points.
    The error bars in both the $x$ and $y$-axis directions denote the corresponding $1\sigma$ uncertainties.
    The grey dashed line represents 1:1 relation.
    }
    \label{fig:rival}
\end{figure}

Here we compare our stellar mass estimates with those from previous studies which employed different techniques on the same sample.
Our sample consists of 24 AGNs, including 6 that also appear in the sample studied by \citet{zou2019}, 4 in the sample studied by \citet{suh2020}, and 6 in the sample studied by \citet{ding2020} (hereafter Z19, S20, and D20).
These overlapping sources are illustrated in Figure \ref{fig:rival}, with the $y$-axis representing our estimates of stellar mass and the $x$-axis denoting those from the corresponding comparison samples.
The derived stellar masses of our sample are presented in Table \ref{tab:sed}. 

Z19 and S20 used similar methods to address the degeneracy between AGN and stellar components in the UV-optical bands.
By assuming energy conservation for galaxies, they constrained stellar emission with FIR photometry.
Compared with Z19, who employed the \texttt{CIGALE} \citep{noll2009, boquien2018} code in the SED fitting, the median absolute difference between our results and theirs is 0.27 dex.
We derived consistent stellar masses for 4 sources with $\log M_\star(\rm M_\odot) > 10.6$.
However, for the other 2 sources that have $\log M_\star(\rm M_\odot)\sim 10.5$ reported by Z19, we derived slightly lower values.
For three of four overlapping sources, our results are inconsistent with S20, who utilized a custom SED fitting code.
We note that our estimates of stellar mass tend to be lower compared to previous SED decomposition results, which could be caused by the additional constraints of image decomposition fluxes that we introduced in our method.

Our estimation is basically consistent with D20, who fitted the decomposed SED with stellar templates to obtain the stellar masses, with a median difference of 0.21 dex.
However, when comparing the points with identical $y$-value in Figure \ref{fig:rival}, we find that the estimates of D20 are basically lower than those given by Z19 and S20.
This implies that optical image decomposition and multiband fluxes may provide inconsistent constraints.
Our method combines the advantages of these two methods. Therefore, the parameter space could be better constrained, which will be discussed in Section \ref{sec:constrain}.

\begin{deluxetable*}{cccccccc}
\caption{Source Catalog} \label{tab:sed}
\tabletypesize{\footnotesize}
\tablehead{\colhead{Name} & \colhead{RA} & \colhead{DEC} & \colhead{$z$} & \colhead{$\log L_{\rm bol}$} & \colhead{$\log \lambda$} & \colhead{$\log M_{\rm BH}$} & \colhead{$\log M_\star$}\\ \colhead{ } & \colhead{(deg)} & \colhead{(deg)} & \colhead{ } & \colhead{$\mathrm{erg\,s^{-1}}$} & \colhead{ } & \colhead{$M_\odot$} & \colhead{$M_\odot$}\\
\colhead{(1)} & \colhead{(2)} & \colhead{(3)} & \colhead{(4)} & \colhead{(5)} & \colhead{(6)} & \colhead{(7)} & \colhead{(8)}
}
\startdata
CID 108 & 150.058685 & 2.477427 & 1.257446 & 45.61 & $-1.31$ & $8.81\pm 0.06$ & $9.95_{-0.39}^{+0.34}$ \\
CID 110 & 150.062134 & 2.455013 & 0.728313 & 44.91 & $-1.59$ & $8.39\pm 0.09$ & $10.13_{-0.17}^{+0.24}$ \\
CID 128 & 150.230789 & 2.578183 & 1.408853 & 45.96 & $-1.65$ & $9.50\pm 0.04$ & $10.63_{-0.28}^{+0.29}$ \\
CID 192 & 149.663589 & 2.085209 & 1.220132 & 45.44 & $-0.80$ & $8.13\pm 0.04$ & $10.24_{-0.16}^{+0.26}$ \\
CID 1930 & 150.042450 & 2.629209 & 1.566184 & 45.74 & $-1.51$ & $9.14\pm 0.05$ & $11.58_{-0.12}^{+0.17}$ \\
CID 216 & 149.791779 & 1.872894 & 1.565879 & 45.04 & $-0.93$ & $7.85\pm 0.10$ & $10.75_{-0.17}^{+0.18}$ \\
CID 3570 & 149.641083 & 2.107653 & 1.242593 & 45.34 & $-0.66$ & $7.89\pm 0.06$ & $11.01_{-0.17}^{+0.20}$ \\
CID $40^\dag$ & 150.199753 & 2.190866 & 1.534250 & 45.69 & $-1.15$ & $8.73\pm 0.24$ & $10.05_{-0.22}^{+0.38}$ \\
CID 597 & 150.526184 & 2.244959 & 1.272449 & 45.50 & $-0.42$ & $7.81\pm 0.10$ & $10.49_{-0.36}^{+0.36}$ \\
CID 607 & 150.609695 & 2.323107 & 1.294432 & 45.83 & $-0.81$ & $8.53\pm 0.03$ & $10.35_{-0.20}^{+0.26}$ \\
CID 644 & 150.511566 & 2.409609 & 0.984967 & 45.07 & $-1.63$ & $8.59\pm 0.05$ & $10.55_{-0.22}^{+0.26}$ \\
CID 71 & 150.123672 & 2.358294 & 0.728576 & 44.91 & $-1.19$ & $7.98\pm 0.03$ & $10.53_{-0.06}^{+0.05}$ \\
CID $72^\dag$ & 150.091537 & 2.399079 & 2.472042 & 45.96 & $-0.47$ & $8.32\pm 0.67$ & $10.66_{-0.22}^{+0.17}$ \\
CID 79 & 150.173965 & 2.402994 & 0.980806 & 45.07 & $-2.04$ & $9.00\pm 0.05$ & $10.04_{-0.27}^{+0.25}$ \\
CID 87 & 150.133041 & 2.303283 & 1.602544 & 45.36 & $-1.31$ & $8.56\pm 0.37$ & $11.63_{-0.18}^{+0.13}$ \\
ECDFS 379 & 53.112499 & $-$27.684805 & 0.740331 & 45.43 & $-1.99$ & $9.31\pm 0.03$ & $10.83_{-0.16}^{+0.14}$ \\
ECDFS 391 & 53.124916 & $-$27.758333 & 1.221274 & 45.73 & $-1.10$ & $8.72\pm 0.26$ & $10.75_{-0.08}^{+0.09}$ \\
LID 1273 & 150.056473 & 1.627495 & 1.620638 & 45.77 & $-0.90$ & $8.56\pm 0.02$ & $10.98_{-0.20}^{+0.20}$ \\
LID 1820 & 149.703598 & 2.578055 & 1.536443 & 45.77 & $-0.89$ & $8.55\pm 0.06$ & $9.58_{-0.13}^{+0.14}$ \\
LID 360 & 150.125092 & 2.861743 & 1.582554 & 46.03 & $-0.53$ & $8.45\pm 0.08$ & $11.03_{-0.21}^{+0.17}$ \\
SXDS 0328 & 34.271252 & $-$5.274861 & 0.807619 & 44.99 & $-2.09$ & $8.97\pm 0.09$ & $10.66_{-0.23}^{+0.20}$ \\
SXDS 0491 & 34.383335 & $-$5.225667 & 1.342809 & 45.67 & $-0.71$ & $8.27\pm 0.03$ & $11.04_{-0.23}^{+0.20}$ \\
SXDS 0610 & 34.469166 & $-$5.259972 & 0.933160 & 45.09 & $-2.84$ & $9.82\pm 0.31$ & $10.81_{-0.21}^{+0.20}$ \\
SXDS 0735 & 34.557919 & $-$4.878194 & 1.442290 & 45.85 & $-0.70$ & $8.44\pm 0.24$ & $9.49_{-0.39}^{+0.37}$
\enddata
\tabletypesize{\footnotesize}
\tablecomments{$\dag$: Sources with BH properties derived from H$\beta$ line. Column (1): object name. Columns (2) and (3): right ascension and declination of the object. Columns (4) (5) (6) and (7): spectroscopic redshift, bolometric luminosity, eddington ratio and BH mass given by \citep{schulze2018}.
Column (6): median stellar mass derived from Bayesian inference, with lower and upper limits indicating 16th and 84th percentiles. 
A machine readable version of this table is accessible through this \href{https://drive.google.com/file/d/1Tw_Z3emDH3xtBeM-a2CZvFgBVO4VnlKn/view?usp=sharing}{google drive link}.}
\end{deluxetable*}


\subsection{\texorpdfstring{$M_{\rm BH}-M_\star$}. relation}

Based on the derived stellar masses, we further investigated the $M_{\rm BH}-M_\star$ relation for our sample.
Due to the fact that the AGN luminosity is likely to depend on black hole mass to a large extent, the selection effects are different for our high redshift sample and the local anchor \citep{lauer2007, schulze2011}, which means that in the high-redshift sample the AGNs with higher black hole masses are more likely to be selected, biasing the $M_{\rm BH}-M_\star$ relation.
For this reason, we first used Monte Carlo Markov Chain (MCMC) simulation to quantify the bias caused by sample selection as \citet{li2021a} did.
Practically, we created a mock catalog of AGNs based on black hole mass function (BHMF) and Eddington Ratio distribution function (ERDF) which were given in \citet{schulze2015}, then performed a luminosity cut to mimic our bolometric luminosity limited sample, as clarified in Figure \ref{fig:sample}.

We used an evolutionary model which is 
\begin{equation}\label{eq:evo}
    \Delta \log M_{\rm BH}=\gamma \log (1+z),
\end{equation}
where $\Delta \log M_{\rm BH}$ refers to the difference between the observed black hole mass and the one inferred from the stellar mass with the local relation \citep{kormendy2013}, hereafter KH13 relation, while $\gamma$ is the evolution factor.
We define $\sigma_\mu$ as the intrinsic scatter in the BH mass.
In the sampling process, we generated 100,000 mock AGNs for each set of ($\gamma$, $\sigma_{\mu}$) values.
We assigned those AGNs with uniformly distributed redshift in the range of $[0.5,2.5]$, then randomly sampled the black hole mass of each AGN ($M_{\rm BH}^{\rm mock}$) according to the BHMF.
Based on the given redshift and black hole mass of each mock AGN, we then sampled the Eddington ratio to obtain their bolometric luminosity.
We removed mock AGNs that have lower bolometric luminosity than $\log L_{\rm bol} = 44.9$ which was the lower limit of our sample, to make the mock sample resemble the observed sample as much as possible.
Then we assigned the stellar mass ($M_\star^{\rm mock}$) of each mock sample based on the black hole masses and the intrinsic scatter in BH mass $\sigma_{\mu}$, which was added as a random normal variable in the model. 

We used likelihood to describe the similarity between the mock sample and the observed sample,
\begin{tiny}
\begin{align}
\mathcal{L}&=\prod\limits_{i=1}^{n} \mathcal{L}_i\notag\\
&= \prod\limits_{i=1}^{n} \left(\frac{N[(\lvert M_\star^{\rm mock} - M_{\star,i}^{\rm obs} \rvert< \epsilon_\star){\rm\ and\ }(\lvert M_{\rm BH}^{\rm mock} - M_{{\rm BH},i}^{\rm obs}\rvert < \epsilon_{\rm BH})]}{N[\lvert M_{\rm BH}^{\rm mock} - M_{{\rm BH},i}^{\rm obs}\rvert < \epsilon_{\rm BH}]} \right),
\end{align}
\end{tiny}
among which $N[\ \cdot\ ]$ stands for the quantity of mock sample that satisfies the condition in the bracket.
$n$ refers to the sample size.
$\epsilon_\star$ and $\epsilon_{\rm BH}$ were determined from the average 1$\sigma$ estimate error of the observed AGNs, which are 0.29 dex and 0.11 dex, respectively.
An analytic derivation of this likelihood function was given by \citet{schulze2011}.

\begin{figure}[htbp]
    \centering
    \includegraphics[width=\columnwidth]{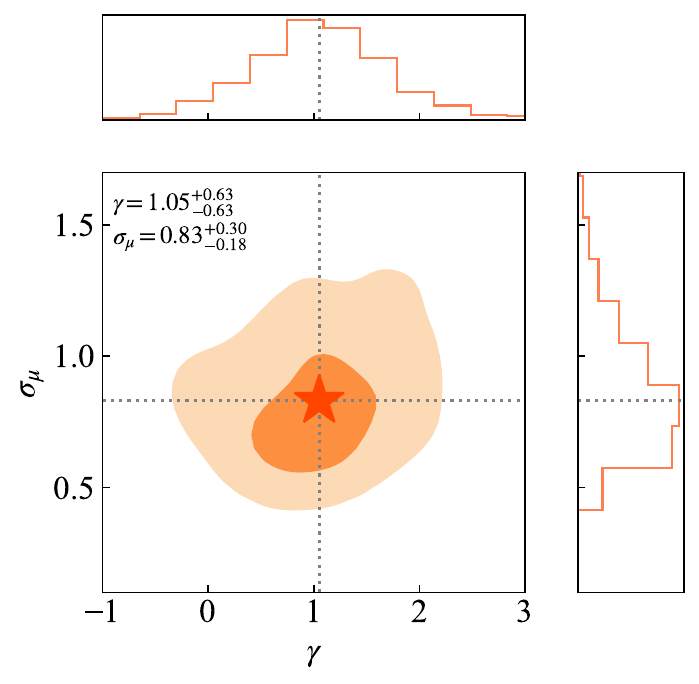}
    \caption{The posterior distributions of the evolution factor $\gamma$ and the intrinsic scatter $\sigma_{\mu}$ of the $M_{\rm BH}-M_\star$ relation, derived from MCMC simulation assuming flat prior on both parameters. The best inference value is denoted as a red star in the central panel. The shaded areas indicate the 68\%, 95\% confidence regions of the posterior probability. The top and right panels show the marginal distributions of the MCMC sample.}
    \label{fig:mcmc}
\end{figure}

Based on the likelihood defined above, we derived the posterior distribution, as shown in Figure \ref{fig:mcmc}, through MCMC inference assuming a flat prior.
Note that some similar studies acquired divergent results under flat prior; thus, they obtained their final result by assuming a log-normal prior on $\sigma_\mu$ in order to constrain the parameter space \citep{ding2020, li2021a}.
Compared with those studies, we derive much higher intrinsic scatter of $\sigma_\mu=0.83_{-0.18}^{+0.30}$ (formerly $\sigma_\mu\sim 0.3$), which may indicate that the black hole and the host galaxy do not grow in lockstep with $z \sim 2$ \citep{yang2018}.

\begin{figure}[htbp]
    \centering
    \includegraphics[width=\columnwidth]{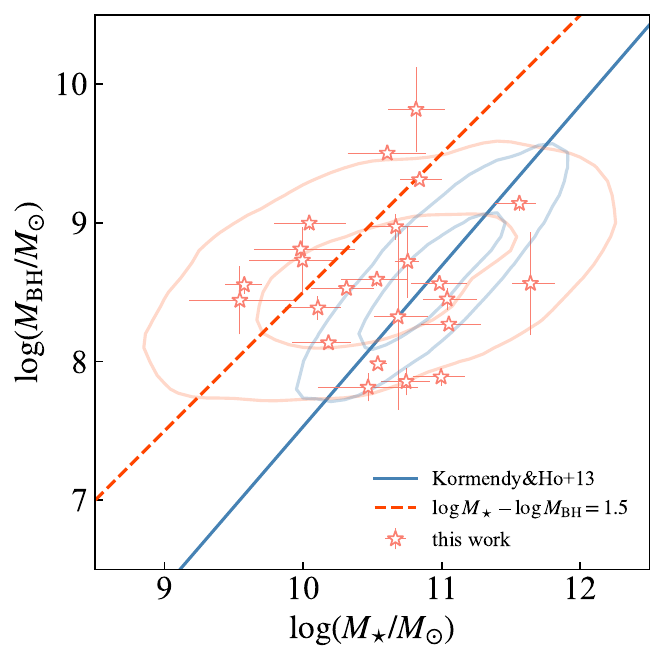}
    \caption{The $M_{\rm BH}-M_\star$ relation of our sample and the comparison sample. 
    The blue line represents KH13 relation.
    The red dashed line represents $\log M_\star - \log M_{\rm BH}=1.5$ ($M_\star$ and $M_{\rm BH}$ in units of $M_\odot$).
    The open orange stars denote our observed sample.
    The error bars represent the $1\sigma$ error.
    The orange contours show 65\% and 95\% confidence region of the mock sample that we reconstructed with $\gamma=1.05$ and $\sigma_\mu=0.83$, while the blue ones represent those reconstructed with KH13 relation parameters of $\gamma=0$ and $\sigma_\mu=0.29$.
    }
    \label{fig:mmrelation}
\end{figure}

The $M_{\rm BH}-M_\star$ relation is plotted in Figure \ref{fig:mmrelation}.
We adopted medians from the Bayesian SED stellar mass estimation, and the error bars represent the interval between the 16th and 84th percentiles of the Bayesian inference sample.
We reconstructed a mock local sample with $\gamma=0$ and $\sigma_\mu=0.29$ and then applied the same luminosity cut as our observed sample to represent the distribution of the mock sample assuming that the correlation does not evolve at all.
Using the model with $\gamma=1.05$ and $\sigma_\mu=0.83$, which is the set of aforementioned best inference parameters, the distribution of the simulated data points shown as orange contours is consistent with the observed ones. 
The orange contours span a wider range compared with the blue ones, which is most likely due to the larger intrinsic scatter.
We note that there are 7 ``BH monsters" lying above the red dashed line, with $\log (M_{\rm BH}/M_\odot)> \log (M_\star/M_\odot) - 1.5$.
If we ignore these ``BH monsters", the other 17 sources lie evenly around the KH13 relation with a median difference in $\log (M_{\rm BH}/M_\odot)$ of 0.43 dex.

\section{Discussion} \label{sec:discussion}

\subsection{Possible sources of error in our work}

Firstly, the multiband fluxes that we use in the SED fitting routine are retrieved from various archival catalogs, which may suffer from inconsistency in photometry.
For example, we adopt optical fluxes from Hubble Souce Catalog \citep{whitmore2016} measured with fixed aperture photometry.
However, in similar optical bands, SDSS catalogs only provide fluxes based on \texttt{modelMag} photometry for extended sources, which were derived based on model fitting of the De vaucouleur profile or exponential profile.
Such inconsistencies may explain the observed fluctuations in data points in the optical bands of the SED (Figure \ref{fig:sed}).
Data from ground-based infrared telescopes such as \textit{VISTA} and \textit{UKIRT} suffer from low resolution and insufficient pointing precision, which is likely to cause the fluxes to include contamination from nearby objects. 

In the AGN-host galaxy image decomposition procedure, the crucial part is PSF selection \citep{ding2020, tanaka2024}. 
The mistake of using extended sources for PSFs could significantly bias the flux decomposition results, particularly given the limited quality of the imaging data used in our work. 
Despite this challenge, we employ a weighted method similar to that proposed by \citet{ding2020}, which utilizes multiple PSFs to mitigate the impact of potential mismatched PSFs. 
Note that \citet{ding2020} collected all available star-like sources around their 32 sources to construct a PSF library.
For each source, they performed an image decomposition using every PSF in the library, ranking the resulting $\chi^2$ and selecting the top N results to cover all possible PSF cases.
However, in five out of ten bands that we used for image decomposition, the number of available images is less than 5, posing challenges in building comprehensive PSF libraries for these bands.
In order to maintain consistency throughout our investigation, we employ an in-between approach for image decomposition, selecting PSFs only from the neighboring regions for each target.
Although this approach may not capture the full variability in PSFs, it enables us to represent the relative goodness of image decomposition for different sources, which is preferable to coarsely fixing all uncertainties at 10\%.
Nevertheless, despite setting a lower limit on the estimation error, mismatched PSFs, if present, could still bias the fluxes of the stellar component towards lower values.

\subsection{Constraining stellar masses} \label{sec:constrain}

\begin{figure*}[htbp]
    \centering
    \includegraphics[width=\textwidth]{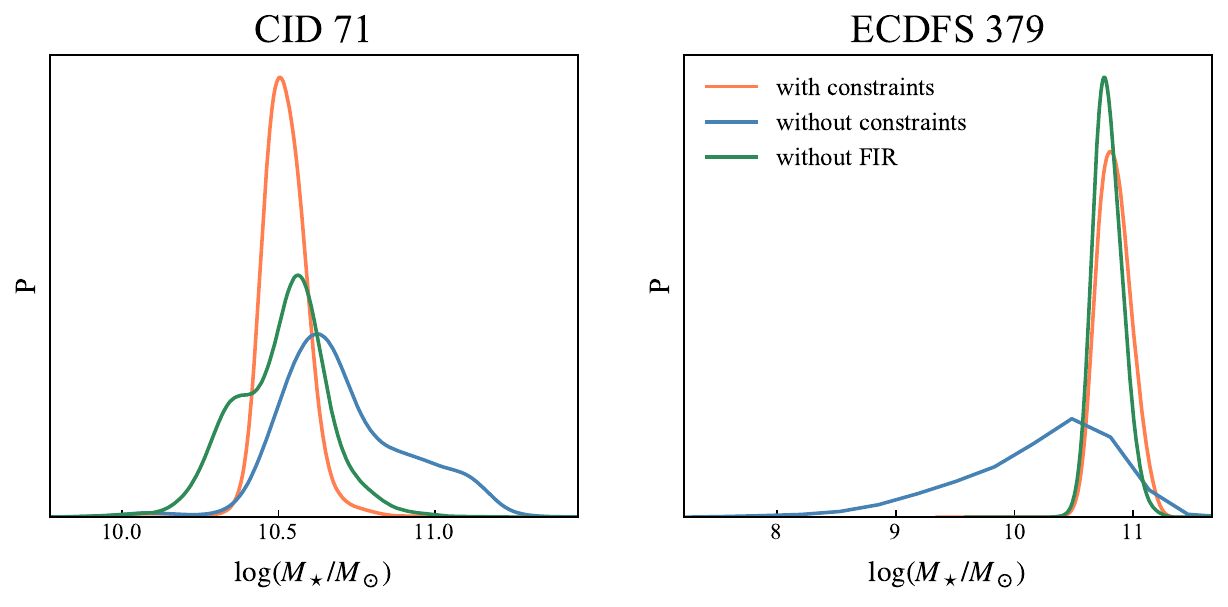}
    \caption{Posterior pdf of $\log (M_\star/M_\odot)$.
    The legend is shown only in the right panel.
    The orange curves represent the derived pdf when employing UV-FIR SED decomposition with additional constraints from image decomposition.
    The blue curves denote the results obtained without these extra constraints (i.e. SED decomposition based solely on broadband photometry).
    The green curves represent the derived pdf using constraints from image decomposition, but excluding FIR data.
    The left and right panels show the results for CID 71 and ECDFS 379, respectively, corresponding to the SED decomposition results shown in Figure \ref{fig:sed}.
    }
    \label{fig:constrain}
\end{figure*}

The primary advancement in our method compared to previous multiband SED decomposition approaches \citep{zou2019, suh2019, suh2020} is the use of image decomposition results as additional constraints to refine SED models.
We examined the posterior distribution of stellar masses with and without these extra constraints, denoted as orange and blue curves in Figure \ref{fig:constrain}.

Although image decomposition could pose strong constraints upon host properties, obtaining high-resolution imaging data for high-redshift samples often requires significant observing time with space telescopes, limiting the sample size in studies on high-redshift AGN host galaxies.
On the other hand, broadband photometric data can be relatively more accessible.
Therefore, we further assessed the relative constraints of broadband photometry in different bands.
Given the considerable uncertainty in FIR observations, we experimented with a two-component model when fitting the data, excluding \textit{Herschel} Pacs and SPIRE FIR data from the input.
The corresponding pdfs are shown as green curves in Figure \ref{fig:constrain}.
For CID 71, excluding FIR photometry significantly increases the estimation error.
However, for ECDFS 379, there is only one photometric point in the FIR bands, so excluding FIR does not make any difference in the results.
As detailed in Section \ref{sec:sed}, the cold dust model is based on the energy balance assumption, thus FIR data provides supplement information about the stellar population. 
We emphasize that the MIR data provide crucial details about the torus \citep{netzer2015} and can put strong constraints on the AGN component, which may help in mitigating the degeneracy between AGN and the stellar component in optical bands.
The optical and NIR fluxes can meanwhile provide supplementary information when the image-decomposed bands are insufficient.

Considering that throughout our study we find no clear bias between our sample with 1 or 2 bands of imaging and those with 3 or more bands, we infer that 2 or 3 bands of imaging is sufficient.
We propose that future work could be based on image decomposition for high-quality imaging data in as few as 3 optical-NIR bands and broadband photometry in other bands, thus expanding the sample size to increase statistical significance.
With \textit{JWST} surveys in NIR-MIR ($4.9-28.1 \rm \upmu m$), investigations on $z\sim 1.5$ type I AGNs with much larger samples would become possible.

\subsection{SFR diagnostics}

We further tested the star formation rates (SFR) obtained from our SED fitting results using the $\mathrm{SFR_{norm}}$ parameter, which quantifies the ``starburstiness'' of an AGN compared to main-sequence (MS) star-forming galaxies.
This parameter is defined as the ratio of the SFR of an AGN and that of MS galaxies of similar $M_\star$ and $z$: $\mathrm{SFR_{norm}}=\frac{\rm SFR_{AGN}}{\rm SFR_{MS}}$.
Following \citet{cristello2024}, who investigated the dependence of $\mathrm{SFR_{norm}}$ on $M_\star$ and $L_{\rm X}$ using \texttt{CIGALE} SED fitting results for AGNs \citep{zou2022}, we analyze the SFR for our samples.

\begin{figure}
    \centering
    \includegraphics[width=\columnwidth]{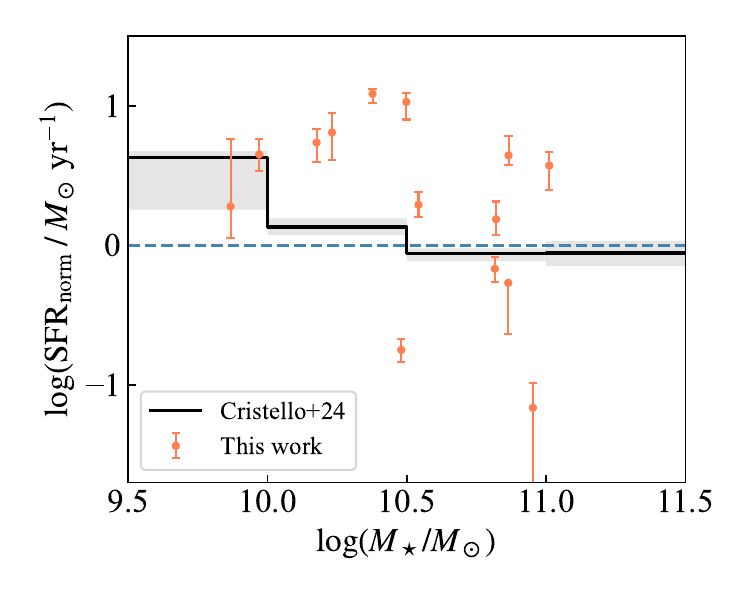}
    \caption{$\log \rm SFR_{norm}$ vs. $\log M_\star$ for sources with FIR photometry in our samples.
    The blue dashed line indicates where the SFR of the AGN host galaxy is identical to that of the MS.
    The black line and the grey shaded area represents the $\rm SFR_{norm} - M_\star$ relation and corresponding 95\% confidence intervals given by \citet{cristello2024}, with $44<\log L_{\rm X} ({\rm erg} ~ {\rm s^{-1}}) < 44.5$.
    The orange points with errorbars represents samples with FIR photometry in this work.}
    \label{fig:sfr}
\end{figure}

Considering that FIR data are crucial for accurately constraining SFR, samples without FIR photometry were excluded in this analysis.
We used the MS data provided by \citet{cristello2024}, which was derived based on a specific star formation rate (sSFR) threshold method.
Specifically, we adopted the MS at $z=1.0$ to approximately match the redshift distribution of our samples.
Since their SFR estimates are derived using similar SED fitting procedures, the adopted MS should be consistent with our sample.
In this subsection, we employ a delayed SFH and a non-evolving chemical abundance model to estimate SFR, which is consistent with \citet{zou2022}. 
Figure \ref{fig:sfr} presents the $ \rm SFR_{\rm norm}$ of our sample, together with the ${\rm SFR_{norm}}-M_\star$ relation for high luminosity ($44<\log L_{\rm X} ({\rm erg} ~ {\rm s^{-1}}) <44.5$) AGNs provided by \citet{cristello2024}.
This range roughly corresponds to the luminosity of our presented sample, which spans from 43.5 to 44.5.
We found that host galaxies with $\log M_\star (\rm M_\odot)<10.5 $  tend to exhibit higher SFR than their MS counterparts,
whereas more massive galaxies with $\log M_\star (\rm M_\odot)>10.5$ show a mean $\rm SFR_{norm}$ approximately equal to unity.
This trend, where less massive AGN hosts tend to exhibit more enhanced star formation, has also been reported by \citet{cristello2024}.
However, our data show a larger scatter compared with the literature, which could be due to the limited sample size.
It is worth noting that the delayed SFH was applied only for the SFR estimation, while a double-power law SFH was adopted throughout the rest of this work.
When comparing these two SFH models, we found no systemic bias in the derived stellar masses.
However, the double power law SFH provided better agreement with other works in Section \ref{subsec:mass}.

\section{Summary} \label{sec:summary}

In this work, we proposed a novel method that combines image decomposition and SED decomposition to constrain stellar masses.
This method enables us to take advantage of limited high-resolution imaging data and mitigate the degeneracy between AGN bright UV-optical emission and host stellar emission as much as possible, constraining host properties in multiple ways.
To further test the viability of this method, we obtained a sample of 24 type I AGNs selected by X-ray observations, with redshift ranging from 0.73 to 2.47. 

The main findings of this work are listed below.
\begin{itemize}
    \item We compared our stellar mass estimates with previous works which have samples overlapping with ours. We find that our results are partially in agreement with those based on the broadband SED decomposition \citep{zou2019, suh2020} and are generally consistent with those based on the image decomposition \citep{ding2020}.

    \item We studied the $M_{\rm BH}-M_\star$ relation of our sample, finding much higher intrinsic scatter for our sample compared to the local $M_{\rm BH}-M_{\rm bulge}$ relation, which may be caused by 7 ``BH monsters" in our sample. If these ``BH monsters" are excluded, our results basically agree with the local correlation given in \citet{kormendy2013}.
    
    \item We evaluated the impact of various data on the derived posterior pdf of stellar masses. The results show that stellar mass is better constrained when using our method, which combines image decomposition data and MIR/FIR broadband photometric data.

    \item We estimated the SFR for samples with FIR photometry and found that less massive AGN hosts ($\log M_\star(\rm M_\odot)<10.5$) tend to exhibit higher SFR compared to MS galaxies, which is consistent with recent studies.

\end{itemize}

We propose that, based on our method which combines image and SED decomposition, future work could rely on as few as 2 or 3 optical-NIR bands for image decomposition and utilize fluxes measured from broadband photometry in any available optical-FIR bands.
Our approach would alleviate the demand for high image resolution and facilitate investigations of larger samples, thereby enhancing our understanding of high redshift AGN host galaxies.

\begin{acknowledgments}

The authors thank the anonymous referee for constructive comments and suggestions. This work is supported by National Key Research and Development Program of China (2023YFA1608100). We gratefully acknowledge the support of the National Natural Science Foundation of China (NSFC, grant No. 12173037, 12233008, 11773063, 12288102, 12025303), the CAS Project for Young Scientists in Basic Research (No. YSBR-092), the China Manned Space Project with NO. CMS-CSST-2021-A02, CMS-CSST-2021-A04, and CMS-CSST-2021-A06, Fundamental Research Funds for Central Universities (WK3440000006) and Cyrus Chun Ying Tang Foundations.
Yunkun Han also gratefully acknowledges the support from the National Key R\&D Program of China (Nos. 2021YFA1600401 and 2021YFA1600400), the ``Light of West China'' Program of Chinese Academy of Sciences, the Yunnan Ten Thousand Talents Plan Young \& Elite Talents Project, the Natural Science Foundation of Yunnan Province (No. 202201BC070003), and the International Centre of Supernovae, Yunnan Key Laboratory (No. 202302AN360001).

\end{acknowledgments}

\vspace{5mm}


\software{\texttt{astropy} \citep{astropy2013, astropy2018, astropy2022}, \texttt{BayeSED} \citep{han2014,han2018,han2023}, \texttt{emcee} \citep{emcee}, \texttt{Galight} \citep{ding2021}, \texttt{getdist} \citep{Lewis:2019xzd}}



\appendix

\section{Image decomposition results}

\begin{deluxetable*}{ccccccccccc}[htbp]
\tabletypesize{\tiny}
\caption{Decomposed host galaxy flux} \label{tab:host}
\tablehead{ \colhead{} & \colhead{} & \colhead{} & \colhead{} & \colhead{} & \colhead{} & \colhead{} & \colhead{} & \colhead{} & \colhead{} \\
\colhead{Source} & \colhead{F435W}& \colhead{F606W}& \colhead{F775W}& \colhead{F814W}& \colhead{F850LP}& \colhead{F105W}& \colhead{F110W}& \colhead{F125W}& \colhead{F140W}& \colhead{F160W}\\
\colhead{} & \colhead{$\upmu \rm Jy$} & \colhead{$\upmu \rm Jy$}& \colhead{$\upmu \rm Jy$}& \colhead{$\upmu \rm Jy$}& \colhead{$\upmu \rm Jy$}& \colhead{$\upmu \rm Jy$}& \colhead{$\upmu \rm Jy$}& \colhead{$\upmu \rm Jy$}& \colhead{$\upmu \rm Jy$}& \colhead{$\upmu \rm Jy$}}
\startdata
CID 108 &$\cdot\cdot$  &$<3.32$  &$\cdot\cdot$  &$10.27\pm 1.91$  &$\cdot\cdot$  &$\cdot\cdot$  &$\cdot\cdot$  &$12.87\pm 1.74$  &$\cdot\cdot$  &$16.88\pm 1.69$           \\
CID 110 &$\cdot\cdot$  &$<2.98$  &$\cdot\cdot$  &$11.35\pm 0.37$  &$\cdot\cdot$  &$\cdot\cdot$  &$\cdot\cdot$  &$21.22\pm 0.44$  &$\cdot\cdot$  &$26.71\pm 0.94$           \\
CID 128 &$\cdot\cdot$  &$\cdot\cdot$  &$\cdot\cdot$  &$1.77\pm 0.59$  &$\cdot\cdot$  &$7.32\pm 2.81$  &$\cdot\cdot$  &$10.44\pm 0.99$  &$\cdot\cdot$  &$9.34\pm 3.79$           \\
CID 192 &$\cdot\cdot$  &$\cdot\cdot$  &$\cdot\cdot$  &$4.48\pm 0.11$  &$\cdot\cdot$  &$\cdot\cdot$  &$\cdot\cdot$  &$\cdot\cdot$  &$\cdot\cdot$  &$12.70\pm 0.21$           \\
CID 1930 &$\cdot\cdot$  &$\cdot\cdot$  &$\cdot\cdot$  &$0.48\pm 0.46$  &$\cdot\cdot$  &$\cdot\cdot$  &$\cdot\cdot$  &$\cdot\cdot$  &$\cdot\cdot$  &$20.82\pm 1.26$           \\
CID 216 &$\cdot\cdot$  &$\cdot\cdot$  &$\cdot\cdot$  &$1.41\pm 0.05$  &$\cdot\cdot$  &$\cdot\cdot$  &$\cdot\cdot$  &$\cdot\cdot$  &$7.03\pm 1.42$  &$\cdot\cdot$           \\
CID 3570 &$\cdot\cdot$  &$\cdot\cdot$  &$\cdot\cdot$  &$\cdot\cdot$  &$\cdot\cdot$  &$\cdot\cdot$  &$\cdot\cdot$  &$11.99\pm 0.42$  &$\cdot\cdot$  &$\cdot\cdot$           \\
CID 40 &$\cdot\cdot$  &$3.53\pm 1.58$  &$\cdot\cdot$  &$4.39\pm 1.85$  &$\cdot\cdot$  &$\cdot\cdot$  &$\cdot\cdot$  &$8.87\pm 1.75$  &$\cdot\cdot$  &$17.74\pm 1.63$           \\
CID 597 &$\cdot\cdot$  &$\cdot\cdot$  &$\cdot\cdot$  &$0.57\pm 0.08$  &$\cdot\cdot$  &$\cdot\cdot$  &$\cdot\cdot$  &$\cdot\cdot$  &$\cdot\cdot$  &$\cdot\cdot$           \\
CID 607 &$\cdot\cdot$  &$\cdot\cdot$  &$\cdot\cdot$  &$0.94\pm 0.09$  &$\cdot\cdot$  &$\cdot\cdot$  &$\cdot\cdot$  &$3.20\pm 0.27$  &$\cdot\cdot$  &$\cdot\cdot$           \\
CID 644 &$\cdot\cdot$  &$\cdot\cdot$  &$\cdot\cdot$  &$1.66\pm 0.17$  &$\cdot\cdot$  &$\cdot\cdot$  &$\cdot\cdot$  &$\cdot\cdot$  &$\cdot\cdot$  &$12.94\pm 0.54$           \\
CID 71 &$\cdot\cdot$  &$7.57\pm 0.12$  &$\cdot\cdot$  &$23.92\pm 0.29$  &$\cdot\cdot$  &$\cdot\cdot$  &$\cdot\cdot$  &$48.64\pm 1.07$  &$\cdot\cdot$  &$63.59\pm 0.50$           \\
CID 72 &$\cdot\cdot$  &$0.36\pm 0.13$  &$\cdot\cdot$  &$1.13\pm 0.90$  &$\cdot\cdot$  &$\cdot\cdot$  &$\cdot\cdot$  &$0.80\pm 0.74$  &$2.44\pm 0.42$  &$3.40\pm 0.27$           \\
CID 79 &$\cdot\cdot$  &$2.13\pm 0.16$  &$\cdot\cdot$  &$2.20\pm 0.03$  &$\cdot\cdot$  &$\cdot\cdot$  &$\cdot\cdot$  &$9.99\pm 0.21$  &$\cdot\cdot$  &$13.24\pm 0.15$           \\
CID 87 &$\cdot\cdot$  &$<0.20$  &$\cdot\cdot$  &$1.24\pm 0.02$  &$\cdot\cdot$  &$\cdot\cdot$  &$\cdot\cdot$  &$10.46\pm 0.10$  &$\cdot\cdot$  &$16.71\pm 0.11$           \\
ECDFS 379 &$2.33\pm 0.18$  &$3.61\pm 0.04$  &$20.31\pm 0.71$  &$21.32\pm 1.02$  &$31.05\pm 3.40$  &$\cdot\cdot$  &$\cdot\cdot$  &$49.83\pm 6.03$  &$47.92\pm 0.14$  &$61.05\pm 3.32$           \\
ECDFS 391 &$0.32\pm 0.04$  &$0.72\pm 0.10$  &$2.08\pm 0.11$  &$2.66\pm 0.02$  &$4.94\pm 0.13$  &$\cdot\cdot$  &$\cdot\cdot$  &$9.46\pm 0.01$  &$13.24\pm 0.22$  &$14.76\pm 0.03$           \\
LID 1273 &$\cdot\cdot$  &$\cdot\cdot$  &$\cdot\cdot$  &$1.30\pm 0.18$  &$\cdot\cdot$  &$\cdot\cdot$  &$\cdot\cdot$  &$\cdot\cdot$  &$8.99\pm 0.12$  &$\cdot\cdot$           \\
LID 1820 &$\cdot\cdot$  &$\cdot\cdot$  &$\cdot\cdot$  &$\cdot\cdot$  &$\cdot\cdot$  &$\cdot\cdot$  &$\cdot\cdot$  &$\cdot\cdot$  &$6.81\pm 1.07$  &$\cdot\cdot$           \\
LID 360 &$\cdot\cdot$  &$\cdot\cdot$  &$\cdot\cdot$  &$0.79\pm 0.47$  &$\cdot\cdot$  &$\cdot\cdot$  &$\cdot\cdot$  &$\cdot\cdot$  &$9.10\pm 1.31$  &$\cdot\cdot$           \\
SXDS 0328 &$\cdot\cdot$  &$\cdot\cdot$  &$\cdot\cdot$  &$\cdot\cdot$  &$\cdot\cdot$  &$\cdot\cdot$  &$\cdot\cdot$  &$11.28\pm 1.11$  &$\cdot\cdot$  &$14.90\pm 2.32$           \\
SXDS 0491 &$\cdot\cdot$  &$\cdot\cdot$  &$\cdot\cdot$  &$\cdot\cdot$  &$\cdot\cdot$  &$\cdot\cdot$  &$\cdot\cdot$  &$10.11\pm 0.50$  &$13.18\pm 0.59$  &$16.81\pm 1.36$           \\
SXDS 0610 &$\cdot\cdot$  &$\cdot\cdot$  &$\cdot\cdot$  &$\cdot\cdot$  &$\cdot\cdot$  &$\cdot\cdot$  &$\cdot\cdot$  &$12.52\pm 0.12$  &$14.27\pm 0.81$  &$18.31\pm 0.68$           \\
SXDS 0735 &$\cdot\cdot$  &$\cdot\cdot$  &$\cdot\cdot$  &$\cdot\cdot$  &$\cdot\cdot$  &$\cdot\cdot$  &$\cdot\cdot$  &$\cdot\cdot$  &$7.02\pm 0.26$  &$\cdot\cdot$           \\
\enddata
\tabletypesize{\footnotesize}
\tablecomments{(a) ``$\cdot\cdot$'' means that data is not available for this band.
(b) ``$<{\rm flux}$'' means that this flux is an upper limit.}
\end{deluxetable*}
\begin{deluxetable*}{ccccccccccc}[htbp]
\tabletypesize{\tiny}
\caption{Decomposed AGN point source flux} \label{tab:ps}
\tablehead{ \colhead{} & \colhead{} & \colhead{} & \colhead{} & \colhead{} & \colhead{} & \colhead{} & \colhead{} & \colhead{} & \colhead{} \\
\colhead{Source} & \colhead{F435W}& \colhead{F606W}& \colhead{F775W}& \colhead{F814W}& \colhead{F850LP}& \colhead{F105W}& \colhead{F110W}& \colhead{F125W}& \colhead{F140W}& \colhead{F160W}\\
\colhead{} & \colhead{$\upmu \rm Jy$} & \colhead{$\upmu \rm Jy$}& \colhead{$\upmu \rm Jy$}& \colhead{$\upmu \rm Jy$}& \colhead{$\upmu \rm Jy$}& \colhead{$\upmu \rm Jy$}& \colhead{$\upmu \rm Jy$}& \colhead{$\upmu \rm Jy$}& \colhead{$\upmu \rm Jy$}& \colhead{$\upmu \rm Jy$}}
\startdata
CID 108 &$\cdot\cdot$  &$23.63\pm 2.36$  &$\cdot\cdot$  &$16.58\pm 2.52$  &$\cdot\cdot$  &$\cdot\cdot$  &$\cdot\cdot$  &$16.98\pm 1.46$  &$\cdot\cdot$  &$24.67\pm 1.86$           \\
CID 110 &$\cdot\cdot$  &$6.36\pm 0.64$  &$\cdot\cdot$  &$5.52\pm 0.08$  &$\cdot\cdot$  &$\cdot\cdot$  &$\cdot\cdot$  &$9.40\pm 0.48$  &$\cdot\cdot$  &$8.46\pm 0.18$           \\
CID 128 &$\cdot\cdot$  &$\cdot\cdot$  &$\cdot\cdot$  &$34.46\pm 0.64$  &$\cdot\cdot$  &$39.38\pm 2.30$  &$\cdot\cdot$  &$43.04\pm 0.70$  &$\cdot\cdot$  &$60.68\pm 3.44$           \\
CID 192 &$\cdot\cdot$  &$\cdot\cdot$  &$\cdot\cdot$  &$17.08\pm 0.21$  &$\cdot\cdot$  &$\cdot\cdot$  &$\cdot\cdot$  &$\cdot\cdot$  &$\cdot\cdot$  &$32.00\pm 0.30$           \\
CID 1930 &$\cdot\cdot$  &$\cdot\cdot$  &$\cdot\cdot$  &$29.96\pm 0.39$  &$\cdot\cdot$  &$\cdot\cdot$  &$\cdot\cdot$  &$\cdot\cdot$  &$\cdot\cdot$  &$33.58\pm 0.93$           \\
CID 216 &$\cdot\cdot$  &$\cdot\cdot$  &$\cdot\cdot$  &$3.31\pm 0.06$  &$\cdot\cdot$  &$\cdot\cdot$  &$\cdot\cdot$  &$\cdot\cdot$  &$4.18\pm 1.46$  &$\cdot\cdot$           \\
CID 3570 &$\cdot\cdot$  &$\cdot\cdot$  &$\cdot\cdot$  &$\cdot\cdot$  &$\cdot\cdot$  &$\cdot\cdot$  &$\cdot\cdot$  &$4.35\pm 0.49$  &$\cdot\cdot$  &$\cdot\cdot$           \\
CID 40 &$\cdot\cdot$  &$9.09\pm 1.59$  &$\cdot\cdot$  &$10.11\pm 1.60$  &$\cdot\cdot$  &$\cdot\cdot$  &$\cdot\cdot$  &$12.04\pm 1.44$  &$\cdot\cdot$  &$11.35\pm 1.45$           \\
CID 597 &$\cdot\cdot$  &$\cdot\cdot$  &$\cdot\cdot$  &$11.07\pm 0.12$  &$\cdot\cdot$  &$\cdot\cdot$  &$\cdot\cdot$  &$\cdot\cdot$  &$\cdot\cdot$  &$\cdot\cdot$           \\
CID 607 &$\cdot\cdot$  &$\cdot\cdot$  &$\cdot\cdot$  &$27.94\pm 0.26$  &$\cdot\cdot$  &$\cdot\cdot$  &$\cdot\cdot$  &$25.90\pm 0.28$  &$\cdot\cdot$  &$\cdot\cdot$           \\
CID 644 &$\cdot\cdot$  &$\cdot\cdot$  &$\cdot\cdot$  &$23.40\pm 0.40$  &$\cdot\cdot$  &$\cdot\cdot$  &$\cdot\cdot$  &$\cdot\cdot$  &$\cdot\cdot$  &$11.72\pm 0.55$           \\
CID 71 &$\cdot\cdot$  &$9.84\pm 0.12$  &$\cdot\cdot$  &$8.13\pm 0.21$  &$\cdot\cdot$  &$\cdot\cdot$  &$\cdot\cdot$  &$11.66\pm 1.09$  &$\cdot\cdot$  &$10.33\pm 0.42$           \\
CID 72 &$\cdot\cdot$  &$3.62\pm 0.15$  &$\cdot\cdot$  &$5.08\pm 0.87$  &$\cdot\cdot$  &$\cdot\cdot$  &$\cdot\cdot$  &$11.54\pm 0.77$  &$10.21\pm 0.42$  &$10.53\pm 0.29$           \\
CID 79 &$\cdot\cdot$  &$0.38\pm 0.13$  &$\cdot\cdot$  &$1.08\pm 0.03$  &$\cdot\cdot$  &$\cdot\cdot$  &$\cdot\cdot$  &$1.81\pm 0.25$  &$\cdot\cdot$  &$1.45\pm 0.18$           \\
CID 87 &$\cdot\cdot$  &$0.24\pm 0.02$  &$\cdot\cdot$  &$0.28\pm 0.03$  &$\cdot\cdot$  &$\cdot\cdot$  &$\cdot\cdot$  &$1.91\pm 0.08$  &$\cdot\cdot$  &$2.20\pm 0.05$           \\
ECDFS 379 &$58.82\pm 0.14$  &$67.32\pm 0.41$  &$63.31\pm 0.54$  &$48.16\pm 0.79$  &$63.74\pm 2.64$  &$\cdot\cdot$  &$\cdot\cdot$  &$69.36\pm 4.59$  &$63.81\pm 1.01$  &$69.55\pm 2.63$           \\
ECDFS 391 &$12.75\pm 0.05$  &$11.81\pm 0.00$  &$11.01\pm 0.06$  &$7.14\pm 0.04$  &$9.06\pm 0.10$  &$\cdot\cdot$  &$\cdot\cdot$  &$7.82\pm 0.01$  &$9.70\pm 0.15$  &$11.54\pm 0.02$           \\
LID 1273 &$\cdot\cdot$  &$\cdot\cdot$  &$\cdot\cdot$  &$26.58\pm 0.31$  &$\cdot\cdot$  &$\cdot\cdot$  &$\cdot\cdot$  &$\cdot\cdot$  &$24.73\pm 0.19$  &$\cdot\cdot$           \\
LID 1820 &$\cdot\cdot$  &$\cdot\cdot$  &$\cdot\cdot$  &$\cdot\cdot$  &$\cdot\cdot$  &$\cdot\cdot$  &$\cdot\cdot$  &$\cdot\cdot$  &$11.51\pm 0.91$  &$\cdot\cdot$           \\
LID 360 &$\cdot\cdot$  &$\cdot\cdot$  &$\cdot\cdot$  &$50.23\pm 1.28$  &$\cdot\cdot$  &$\cdot\cdot$  &$\cdot\cdot$  &$\cdot\cdot$  &$47.22\pm 1.26$  &$\cdot\cdot$           \\
SXDS 0328 &$\cdot\cdot$  &$\cdot\cdot$  &$\cdot\cdot$  &$\cdot\cdot$  &$\cdot\cdot$  &$\cdot\cdot$  &$\cdot\cdot$  &$14.76\pm 0.95$  &$\cdot\cdot$  &$13.70\pm 1.98$           \\
SXDS 0491 &$\cdot\cdot$  &$\cdot\cdot$  &$\cdot\cdot$  &$\cdot\cdot$  &$\cdot\cdot$  &$\cdot\cdot$  &$\cdot\cdot$  &$8.69\pm 0.40$  &$9.90\pm 0.64$  &$11.81\pm 0.89$           \\
SXDS 0610 &$\cdot\cdot$  &$\cdot\cdot$  &$\cdot\cdot$  &$\cdot\cdot$  &$\cdot\cdot$  &$\cdot\cdot$  &$\cdot\cdot$  &$7.93\pm 0.16$  &$6.98\pm 1.18$  &$6.96\pm 0.58$           \\
SXDS 0735 &$\cdot\cdot$  &$\cdot\cdot$  &$\cdot\cdot$  &$\cdot\cdot$  &$\cdot\cdot$  &$\cdot\cdot$  &$\cdot\cdot$  &$\cdot\cdot$  &$44.00\pm 0.36$  &$\cdot\cdot$           \\
\enddata
\tabletypesize{\footnotesize}
\tablecomments{``$\cdot\cdot$'' means that data is not available for this band.}
\end{deluxetable*}

\bibliography{main}{}
\bibliographystyle{aasjournal}



\end{document}